# What country, university, or research institute, performed the best on COVID-19 during the first wave of the pandemic?

## Bibliometric analysis of scientific literature – analysing a 'snapshot in time' of the first wave of COVID-19

**Petar Radanliev[1] · David De Roure[1] · Rob Walton[1] · Max Van Kleek[2] · Omar Santos[3] · La'Treall Maddox[3]**

**Abstract**
In this article, we conduct data mining and statistical analysis on the most effective countries, universities, and companies, based on their output (e.g., produced or collaborated) on COVID-19 during the first wave of the pandemic. Hence, the focus of this article is on the first wave of the pandemic. While in later stages of the pandemic, US and UK performed best in terms of vaccine production, the focus in this article is on the initial few months of the pandemic. The article presents findings from our analysing of all available records on COVID-19 from the Web of Science Core Collection. The results are compared with all available data records on pandemics and epidemics from 1900 to 2020. This has created interesting findings that are presented in the article with visualisation tools.

## 1 Introduction

Since the COVID-19 pandemic started, we have seen an increasing number of scientific research articles, on a wide variety of related topics to disease, pandemics,

✉ Petar Radanliev
petar.radanliev@eng.ox.ac.uk; petar.radanliev@oerc.ox.ac.uk

1 Department of Engineering Sciences, University of Oxford, Oxford, England, UK

2 Department of Computer Science, University of Oxford, Oxford, England, UK

3 Cisco Research Centre, Research Triangle Park, North Carolina, USA

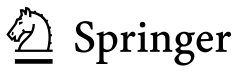

viruses, etc. Some of these topics are closely related to technological advancements and data sciences solutions. For example, the research on tracking and monitoring of cases, is closely related to digital solutions, e.g. mobile apps.

In this study, we use Web of Science data records from a 'snapshot in time' (published until 16th of May 2020) with a computable statistical method, to investigate the correlations between, different scientific research records on the COVID-19 pandemic. Apart from investigating the connections between different topics, we also investigate the data records for **patterns** in the response from different countries. Our objective is that by investigating individual responses, we can provide scientific insights on specific organisations performances, e.g. the World Health Organisation (WHO) speed of response. There are topics that we consider beyond the scope of this study, such as the concerns on the origin of the disease. Our aim is to provide statistical analysis, that can assist other researchers in answering these difficult questions.

With the global focus on the pandemic, the data records are changing dramatically. Since research data records are often categorised by year and not by months, it could be challenging for researchers to find scientific data and to model, with precision, the research response at different stages of the pandemic. We considered this study to be of significant relevance because it provides statistical results that can be seen as a snapshot in time. Our rationale was based on the fact that at the time of the 'snapshot' the pandemic had been in existence for a few of months, and the scientific per-review process last few months. Hence, the data that we collected during the 'snapshot', is from research papers that have been produced at the very beginning of the pandemic.

We applied semi-automatic and automatic analysis of big data, to extract unusual and unknown patterns, from data records on COVID-19, published until **16th of May 2020**. We analysed 3094 data records, which constituted all data records in existence **at the time of the snapshot** - from the Web of Science Core Collection on COVID-19. To compliment this, we conducted a second analysis of 138,624 historical data records from the Web of Science Core Collection, on pandemics and epidemics, covering the time period from 1900 to 2020. We used the historical data records, to compare with the current scientific research on COVID-19, and we use quantitative analysis to derive unexpected conclusions on the speed of response, from the most prominent organisations in pandemic research. In the investigation, we applied cluster analysis, anomaly detection analysis, association rule mining, and sequential pattern mining, among other methods. The findings of this study are presented in groupings of data records, and categorisations of patterns from the input data, which can be used or reproduced in future studies for predictive analytics, e.g. forecasting, monitoring and management of future pandemics.

### 1.1 Research questions

Our objectives are to use computable statistical methods, to conduct bibliometric data mining on scientific research records and to answer some emerging questions on COVID-19. In the study, we investigate:

1. What country produced the most research papers on Covid-19 since the pandemic started?

2. What universities and companies are publishing research on Covid-19?
3. Which countries/universities collaborated most in research papers on Covid-19?

After identifying the answers to these research questions, we focus on a new set of research questions:

4. What country produced the most research papers on pandemic and epidemics from 1900 to 2020?
5. What universities and companies have published most research on pandemic and epidemics from 1900 to 2020?
6. Which countries/universities collaborated most in research papers on pandemic and epidemics from 1900 to 2020?

We use a variety of statistical methods (e.g. three-fields plot, factorial analyse, historical analysis, network map analysis, etc.) to compare the findings from these questions.

### 1.2 Discussion on data science

Data science consists of (1) design for data; (2) collection of data; and (3) analysis on data; and can be described as 'synthetic concept to unify statistics, data analysis and their related methods' [1]. Data science practitioners apply integrated techniques to analyse real-world big data problems. Some of the integration concepts of big data analytics and/or data science include 'multi-criteria optimization for learning, expert and rule-based data analysis, support vector machines for classification, feature selection, data stream analysis, learning analysis, sentiment analysis, link analysis, and evaluation analysis' [2]. Data mining practical applications in various fields (e.g., financial analysis, credit management, health insurance, network intrusion detection, internet services analysis) are often enhanced with optimisation techniques, such as (1) Support Vector Machines for Classification; (2) LOO Bounds for Support Vector Machines; (3) Unsupervised and Semi-supervised Support Vector Machines; (4) Feature Selection via $l_p$ - Norm Support Vector Machines; (5) Multiple Criteria Programming; etc. [3]. Data science is strongly represented in business data mining [4], for real-time decision making with a combination of internet-of-things (IoT) and artificial intelligence (AI) technologies [5]. More recently, the Covid-19 pandemic has been analysed with various data science tools, e.g., 1 for outbreak prediction of the top 10 highly and densely populated countries, using Auto-Regressive Moving Average [6]; e.g., 2 for 'age-specific social contact characterization. of the underlying transmission patterns' [7]. The data mining in this article is more closely related to 'culture vs. policy' [8] with the aim of promoting more global collaborations to combatting global pandemics with technological solutions.

#### 1.2.1 Data mining vs. data analysis

In this study, we differentiate our data mining approach from data analysis. We consider data analysis to be related to testing the effectiveness of specific models or hypothesis. We differentiate this from the data mining in our study, which we consider as using computerised statistical models to uncover interesting, unusual and unknown patterns from big data. Therefore, any reference to analysis in this study, e.g. historical analysis, bibliometric analysis, etc., refers to data mining and not data analysis.

In addition, we understand that our data mining is based on keywords which were representative of the pandemic in the 'snapshot in time' that we analysed. With time, these keywords will change and evolve, and the characteristics of the future analysis should also evolve with the characteristics of the new data records. Nevertheless, this evolution will happen in the future, and with this article, we wanted to preserve the information as it was during the 'snapshot in time' which was taken during the first wave of the Covid-19 pandemic – snapshot was taken in May 2020 and represents the time period from December 2019 when Covid-19 emerged, to May 2020 when the first wave ended – although, there might be various different interpretations as to the exact end date of the first wave.

## 2 Literature review

We conducted a brief literature review, to identify the current gaps in knowledge and to structure our research questions around these gaps. We found a related study on scientometric analysis of COVID-19 and coronaviruses [9]. Hence, we structured our questions on bibliometric analysis of COVID-19, compared with historic data on pandemics and epidemics. The main differences in this article in the approach. Scientometric analysis is focused on the performance of different authors, or journals. The bibliometric analysis in this article is focused on analysing national responses, institutional outputs, and the correlations between research findings. Similar research is present from March 2020 [10], and presents analysis of 564 data records. Since then, the number of scientific research data records has increased to 3094. In addition, we use different statistical methods in our data mining and visualisation, which enables us to compare the COVID-19 analysis, with 138,624 historical data records on pandemics and epidemics. This is significantly different that a bibliometric analysis of 564 data records. The third study we reviewed to structure our research questions, was a study from March 2020, based on 183 data records from PubMed and analysed Identified and analysed the title, author, language, time, type and focus [11]. To differentiate our focus on looking at the same problem, from a different aspect, we used Web of Science data records, and we focused on clustering, classification, association, regression, summarisation, and anomaly detection.

Prior to conducting this review study, we consulted earlier articles on bibliometric analysis and review on artificial intelligence in health care [12], on roles and research trends analysis with bibliometric mapping analysis and systematic review [13], and on the role of bibliometric and review in different research areas e.g., in operations environment [14].

The innovation of the bibliometric analysis in this paper is the categorisations of one research keyword (Covid-19) in a separate analysis with its main research area (pandemics and epidemics). This innovation enables the review to derive with two postulates on what country, university, or research institute, performed the best on COVID-19 during the first wave of Covid-19. The postulates are analysed in great debt with bibliometric analysis of scientific literature from a 'snapshot in time' of the first wave of Covid-19.

## 3 Methodology

In this study, we applied computable methods for statistical analysis, including R Studio, 'Biliometrix' package [15], and VOSviewer [16]. To extract big data from established scientific databases, we used the Web of Science Core Collection, which contains data records from over 21,100 peer-reviewed, high-quality scholarly journals published worldwide, in over 250 disciplines[1].

### 3.1 Data mining on COVID-19

Data mining represents a process of discovering new knowledge from patterns in big data. Usual methods applied include a combination of machine learning and statistics, on analysing big database systems. Data mining is considered a research field that combines computer science and statistics, in designing intelligent methods for extracting new information and for knowledge discovery from existing databases.

The data mining in this study involved data management, data pre-processing, model inference and complexity considerations, postprocessing of discovered results considerations, visualisation, and interestingness metrics.

## 4 Bibliometric analysis

Bibliometric analysis, or bibliometrics, is the practice of using statistical methods to analyse research publications, books, articles, and other scientific communications. In this bibliometric analysis, we extracted data records from the Web of Science Core Collection, and we analysed the records with three different data mining tools, (1) the Web of Science analyse results built-in tool; (2) data mining with VOSviewer; (3) data mining with the R Studio 'Bibliometrix' package.

### 4.1 Data records

The first search for data records was on the Web of Science Core Collection. We searched for all data records on COVID-19 and we extracted 3094 data records (search performed on the date: 17th of May 2020). We also conducted a second search for TOPIC: (pandemic) OR TOPIC: (epidemic), which resulted with 138,624 data records. Both data sets were analysed with the Web of Science analyse results built-in tool. Only the smaller data set was analysed with computerised statistical analysis, using the R Studio 'Bibliometrix' package. This was not by choice, but because of practicality. The Web of Science has data extraction limit of 500 records, to download the 3094 data records, we extracted 7 different files, and we merged the files using the 'Sublime Text' program. To repeat this process on 138,624 data records, we would need to extract 277 separate data files, and merge into one. This was considered tiresome and not practical. Instead, for the VOSviewer data mining, we used the Web of Science tool to identify the top 1000 most relevant data records and we used this data set as representative sample of the 138,624 data records. Only the 1000 most relevant data records are used for the VOSviewer data mining on pandemic and epidemics.

[1] https://clarivate.com/webofsciencegroup/solutions/web-of-science-core-collection/.

### 4.2 Automatic data mining using the web of Science analyse results built-in tool

To analyse all data records available on the Web of Science Core Collection, we used the built-in result analysis tool. First, we categorised the data records in researcher areas Fig. 1.

From Fig. 1, we can see that current research is focused on the medical aspect of COVID-19. There is very little scientific research on the digital aspect of monitoring and managing the pandemic. Other relevant research areas are also missing, such as guidance on privacy preserving mobile app design for pandemic management, the role of internet of things in pandemic management, philosophical perspective on long term societal changes caused by the pandemic. In the first wave of the pandemic, the focus seems to have been predominantly on the medical aspects. Learning from this result, we can conclude that all other research areas become secondary in the immediate threat of pandemics - death. Therefore, scientific research on these topics should be ongoing and constantly advancing, in anticipation of similar pandemics happening in the future, without notice. To analyse if such preparations were happening in the past, we analyse the data records on COVID-19, and we compare the results with a historical analysis of data records on pandemic management. In Fig. 2, we categorise the data records by country.

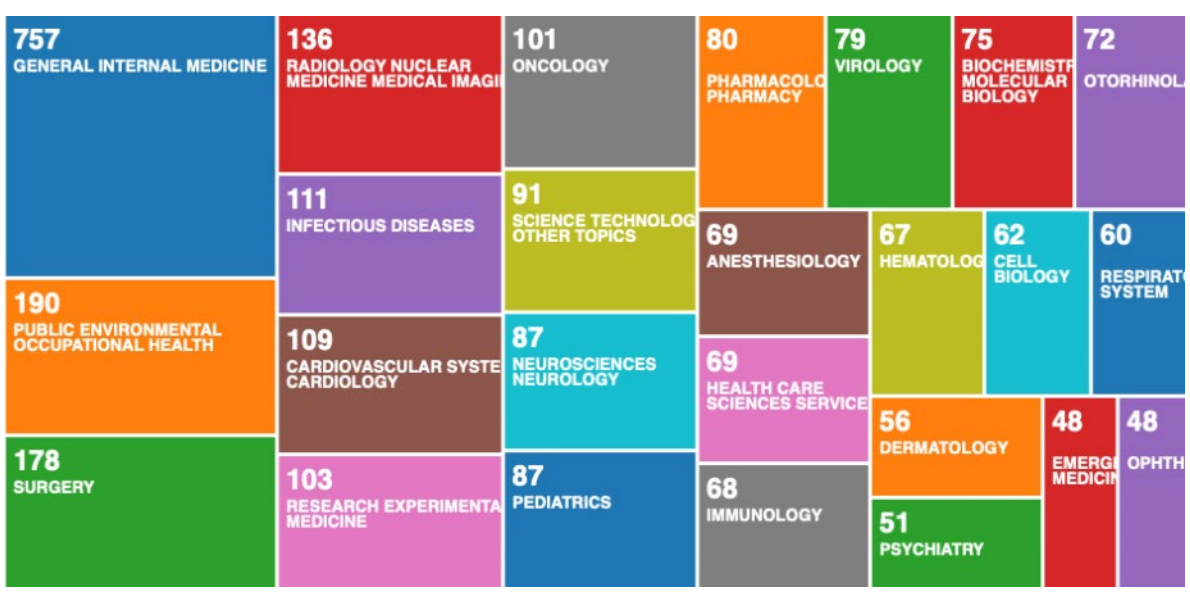

**Fig. 1** Web of Science result analysis tool – research areas

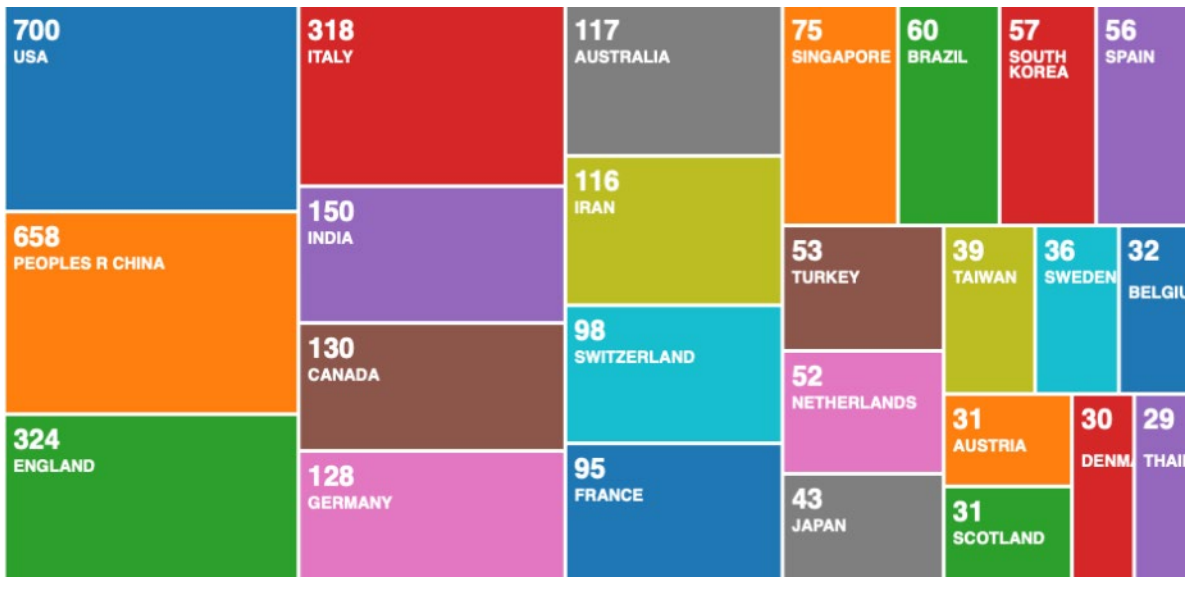

**Fig. 2** Web of Science result analysis tool – research by country

From Fig. 2, we can see that most scientific research is happening in the US and China, followed by the UK and Italy. Would be interesting to compare these results after few weeks, and see if the countries produce more output as the infections spreads. We would suggest focusing on India, because of how the virus spread. With time, we could see the output from India increasing, if we are correct in our assumption that output increases as countries are faced with the deadly pandemic. The leading countries in Fig. 2, are some of the worst affected countries at the time we collected our data records. Although Spain and Iran are also in the hardest hit countries, the scientific research from these two countries is not showing as strong. Therefore, it is indicative, but not conclusive that countries that are most affected, are also most productive in terms of scientific research.

To advance this analysis, we categorised the data records by organisations (enhanced) in Fig. 3. What becomes visible from the categorisation in Fig. 3, is that among the most reputable universities, which usually predominate such categorisations, we now have Wuhan University, where the pandemic originated (was first detected).

The organisations (enhanced) in Fig. 3 categorises the 3094 data records, to include research from associated organisations. We compare the 3094 data records on COVID-19, with the second data file on pandemics and epidemics records from 1900 to 2020, containing 138,624 data records in Fig. 4.

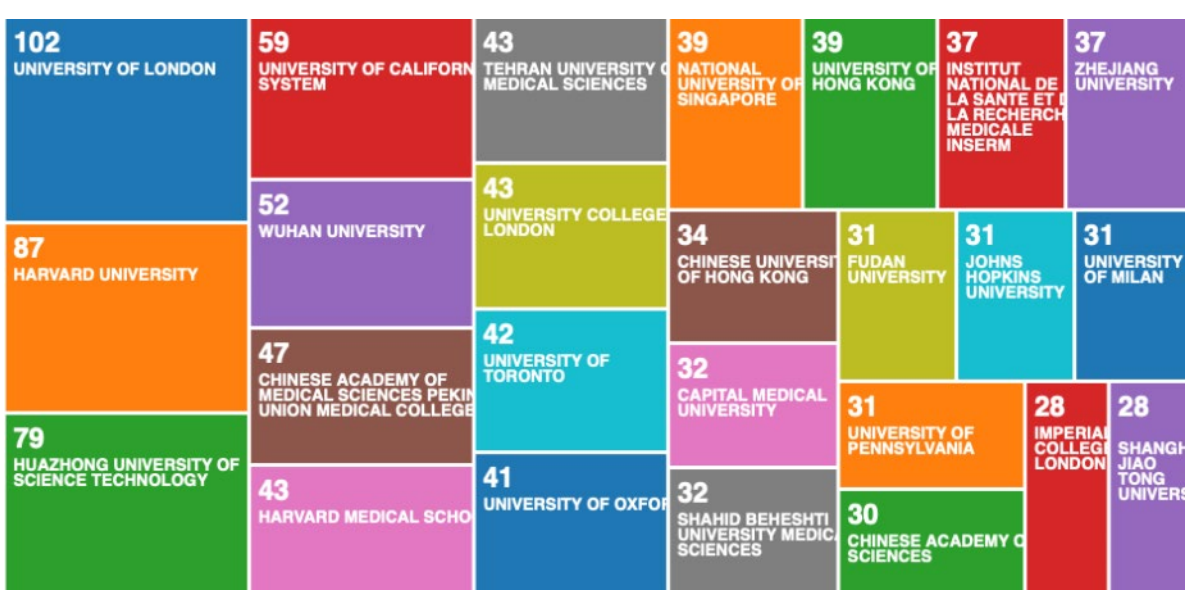

**Fig. 3** Web of Science result analysis tool – research on COVID-19 by organisation (enhanced)

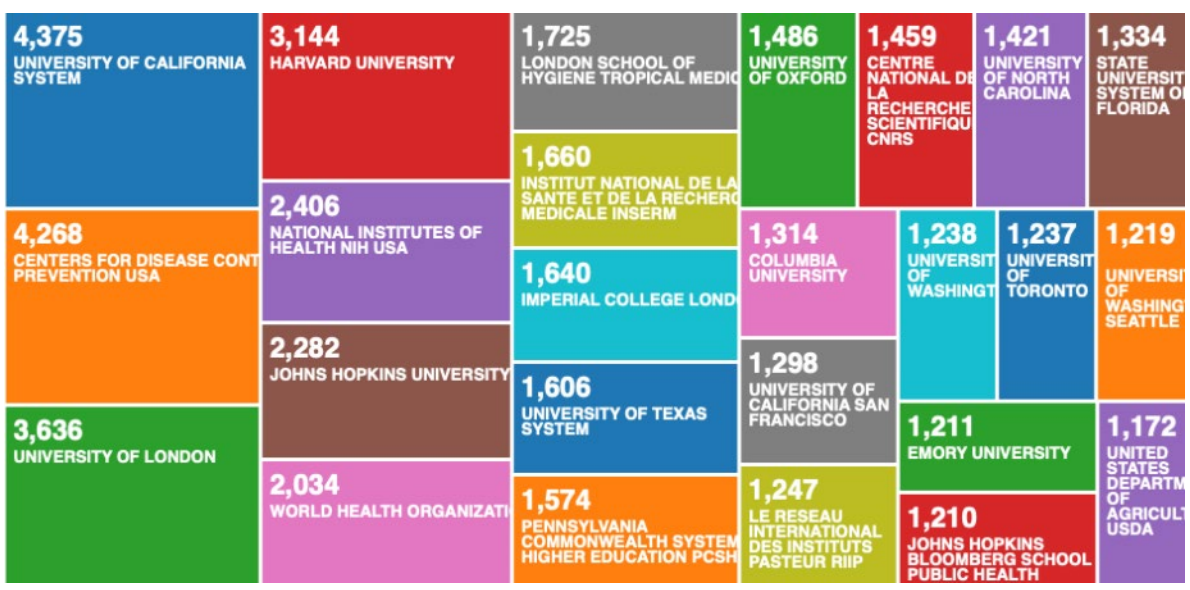

**Fig. 4** Web of Science result analysis tool – research on pandemics and epidemics by organisation (enhanced)

What becomes clear when we compare the two classifications from Figs. 3 and 4, is that some of the best performing universities on COVID-19, are not even present on the list of best performing research organisations on global pandemics and epidemics from the historical analysis. This indicates that there is either a global shift in scientific research, or the early affected regions e.g. Wuhan, have been most productive in scientific research on COVID-19. The second seems more likely.

Since the Figs. 3 and 4 are classifying organisation-enhanced categories, to get a different perspective on organisations own research production, we did a second categorisation of the 3094 data records, by organisations own research Fig. 5.

By categorising the organisations own research, we present a different result from the same data records. In the simple categorisation Fig. 5, we can see that Chines universities are currently in the lead, and we can also see that University of Teheran is also working on this topic, and its much higher ranked in terms of productivity from the previous categorisations - top performing organisation enhanced categorisation in Fig. 3.

When we compare the Fig. 5 - which is visualising the 3094 data record file, with Fig. 6 - which is visualising the 138,624 data record file, we can see a further confirmation that the top performing institutions by output on COVID-19 (Fig. 5), are not representative of the best performing research institutions (Fig. 6). This could signify that the world, for unclear reasons, was slow in responding with scientific research on COVID-19. We could speculate that the world didn't take COVID-19 seriously, or that Chinese knew something that the rest of the world didn't, but we have no data to confirm such speculations. What we can confirm with certainty, is that the Chinese research institutes acted much faster than the rest of the world, including the leading research organisation on pandemics and epidemics.

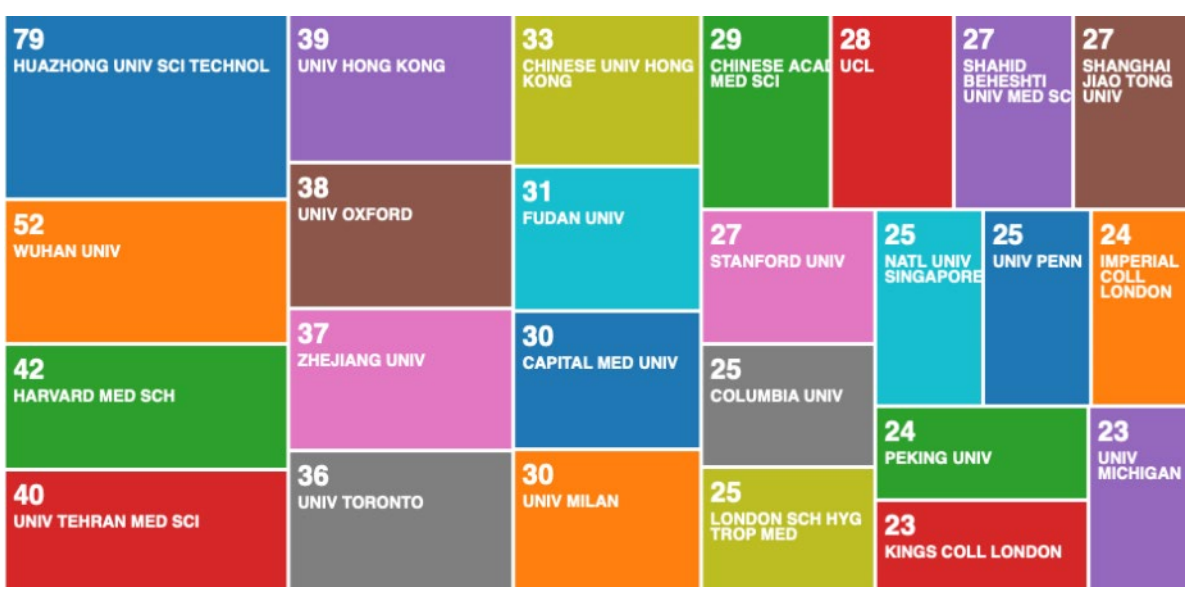

**Fig. 5** Web of Science result analysis tool – research on COVID-19 by organisation (simple categorisation)

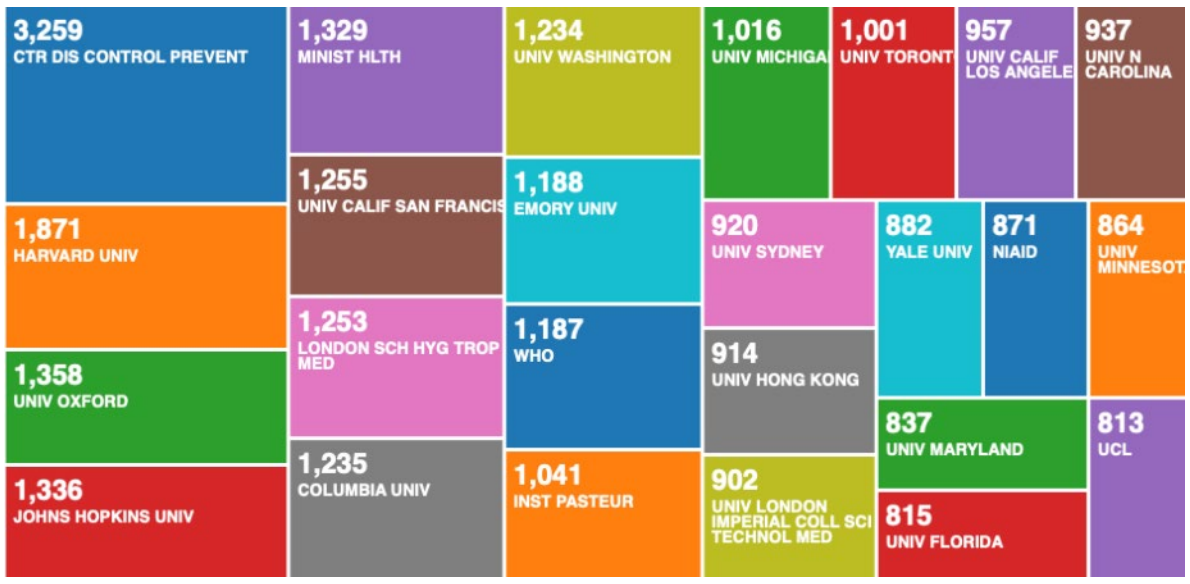

**Fig. 6** Web of Science result analysis tool – research on pandemics and epidemics by organisation (simple categorisation)

Finally, the last categorisation in this part of the analysis, we investigated the scientific research published by funding agencies Fig. 7.

What we can see in the categorisation by funding (based on the 3094 data record file) in Fig. 7, is that China is in the lead, but the US has more distributed funding programme, and if we sum up all the funding, we could get a different result. What is surprising however, is the weak performance of EU funding agencies. There are only 6 data records from the EU funds.

When we compare the Fig. 7 - which is visualising the 3094 data record file, with Fig. 8 - which is visualising the 138,624 data record file, we can see that the organisations that have historically provided most of the funding on pandemics and epidemics, are not in the lead.

Since COVID-19 is a global pandemic, the classifications in Fig. 7, should be similar with Fig. 8. The results are very different, and it is uncertain why the global response was so much slower than the Chinese response. But when we compare the institutions in Fig. 5, with countries that got most affected in the early stages of COVID-19, we can see the connection between countries affected early, and increased data records. Such assumptions from the categorisations, based on the automatic data mining using the Web of Science analyse results built-in tool, are speculative. We need more specific data mining methods to analyse this data records further. In the following section, we apply semi-automated data mining to look for association rule learning, anomaly detection, and regression to accompany and enhance our clustering and classification.

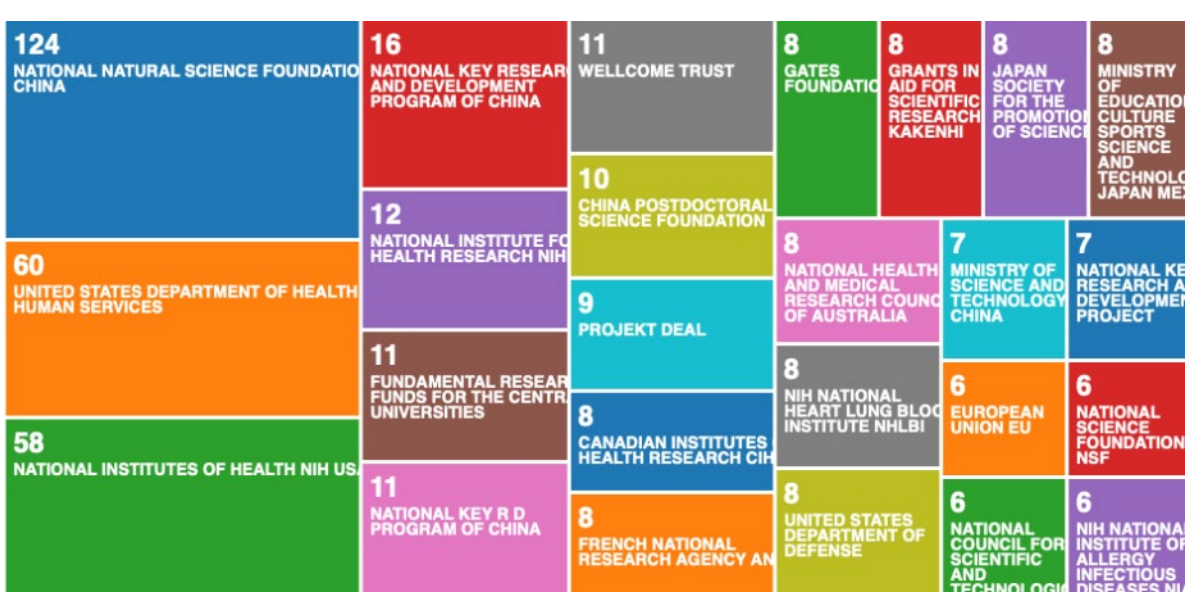

**Fig. 7** Web of Science result analysis tool – research on COVID-19 by funding agencies

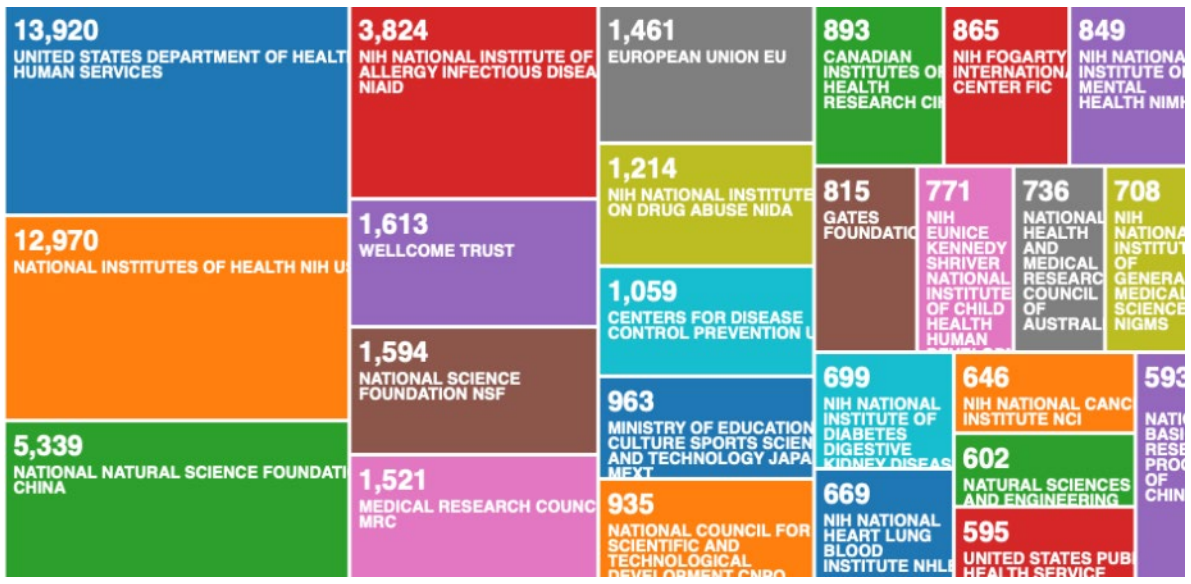

**Fig. 8** Web of Science result analysis tool – research by funding agencies on pandemics and epidemics (historic records)

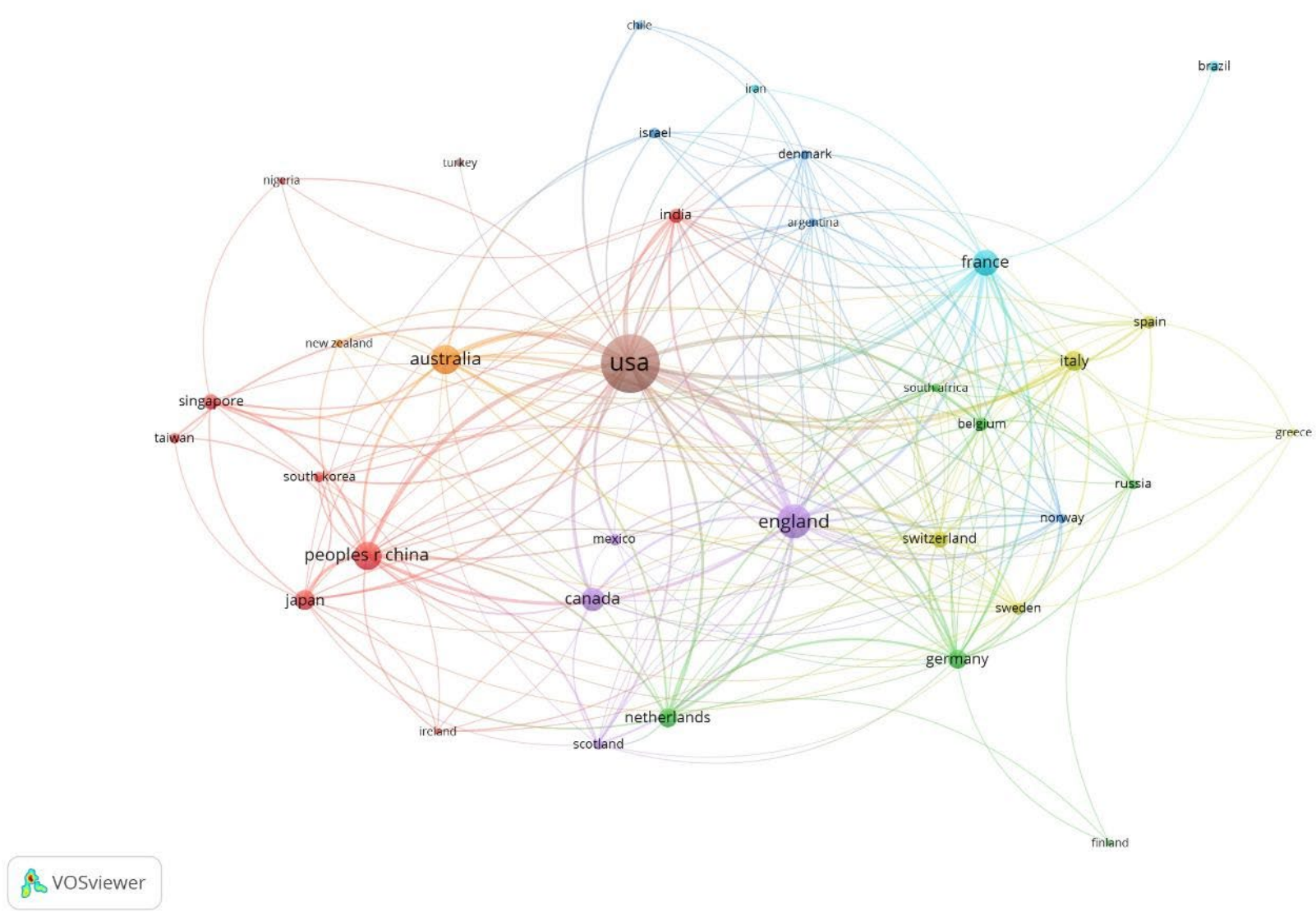

**Fig. 9** Top 1000 most relevant data records on pandemic and epidemic - visualisation by country

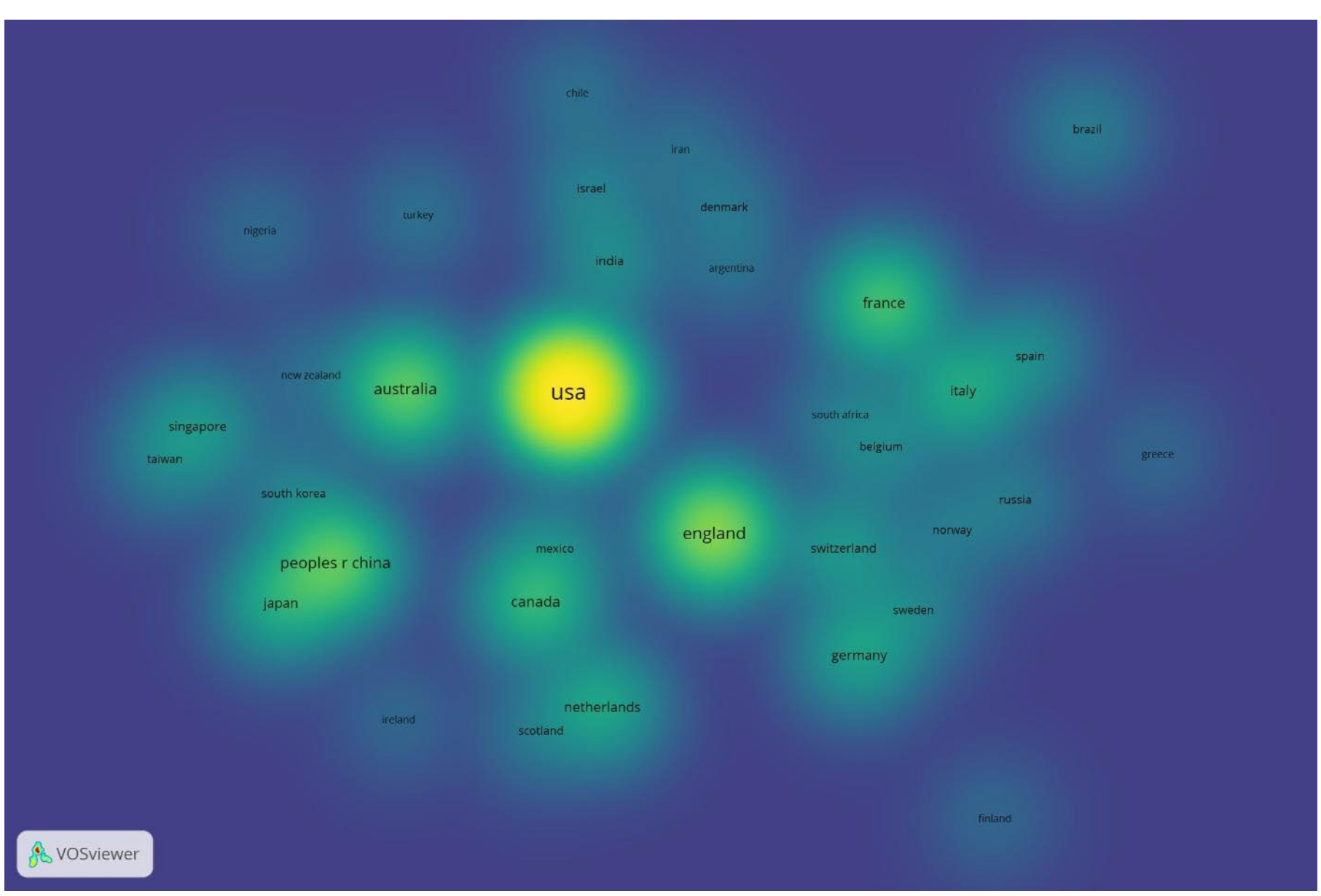

**Fig. 10** Density visualisation by country of the top 1000 most relevant data records on pandemic and epidemic

### 4.3 Semi-automatic data mining with VOSviewer

We continued our data mining with using a computerised statistical analysis, using the VOSviewer computer program. From the 138,624 data records on pandemic and epidemics on the Web of Science Core Collection, for the VOSviewer data mining, we used the top 1000 most relevant data records and we considered this data set as representative sample of the 138,624 data records. We exported two separate text files and we used the two files for the data mining with VOSviewer. In Fig. 9, we can see the VOSviewer visualisation by country and collaborations between countries. In VOSviewer, we can select specific relationships of one country, and we can zoom in the image for more detailed data mining. It is however relatively easy to identify the US, England, Australia and China as the leading countries in the top 1000 historical data records on pandemics and epidemics.

The Figs. 9 and 10 are both based on Lin/Log modularity normalisation. We conducted normalisation by association strength, and by fractionalisation normalisation, but the Lin/Log modularity normalisation presented better visualisation of the research collaborations between countries. For data mining on collaborations between institutions, we used associated strength normalisation, and since we wanted to investigate the collaborations, we set a limit on data records that included collaborations, this reduced our data records from 1000 to 90 analysed in Fig. 11.

Although it's difficult to see in the Fig. 11 visualisation, the VOSviewer identified 8 clusters, with the US universities predominating the biggest two clusters, and Chinese universities appearing in the third cluster. We continued our data mining in the next section with using a computerised statistical analysis, using the R Studio 'Bibliometrix' package.

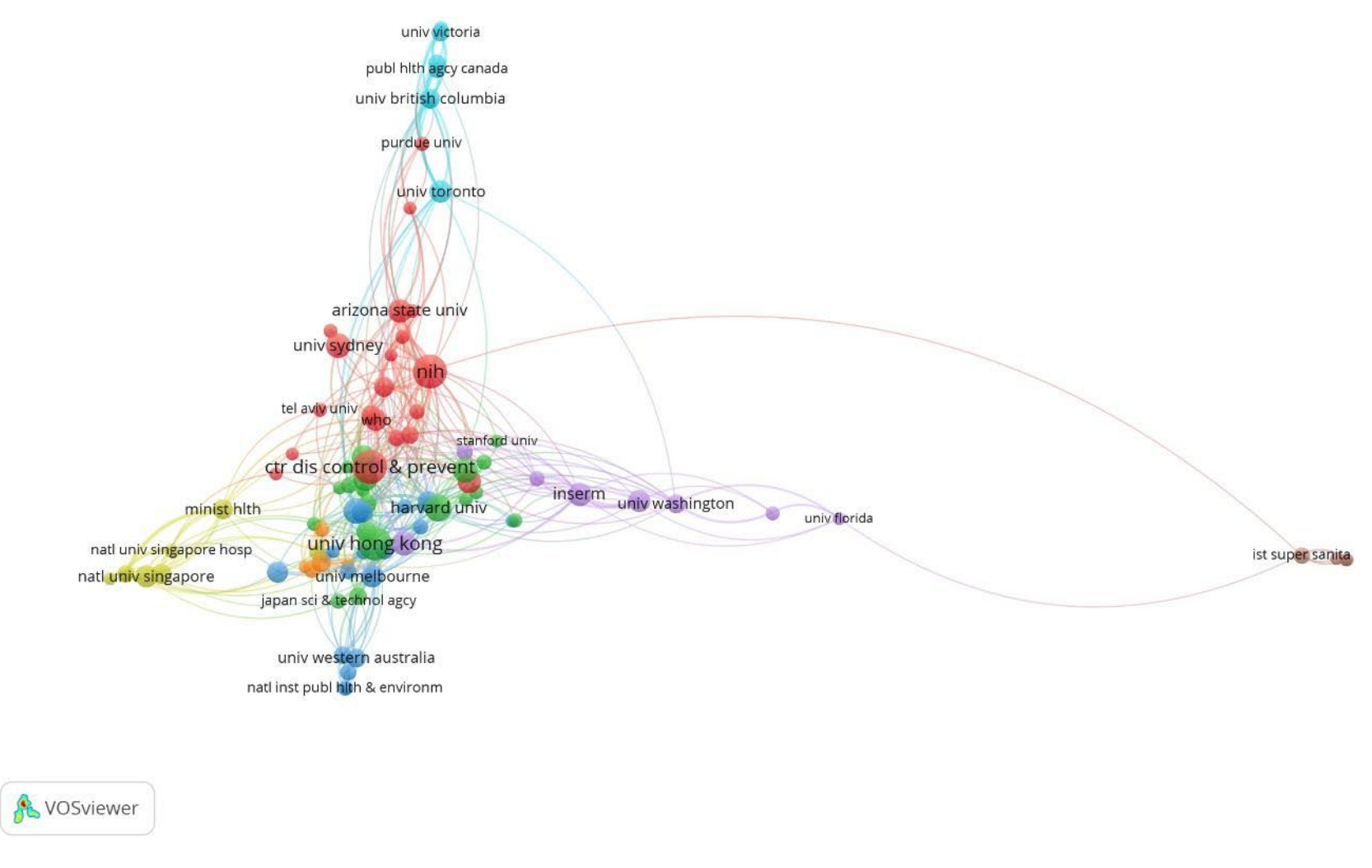

**Fig. 11** Collaborations between institutions on pandemics and epidemics historically and globally - normalisation based on association strength

### 4.4 Semi-automatic data mining with R Studio 'Bibliometrix' package

Since the Web of Science has data extraction limit of 500 records, to download the 3094 data records, we extracted 7 different files, and we merged the files using the 'Sublime Text' program. Then we downloaded the file in the 'Bibliometrix' package, 'Biblioshiny' function. Our data mining was based on association rule and clustering, using thee-fields plots, factorial analysis, collaboration network, conceptual map design, etc.

The first graph we present (Fig. 12) is based on association rule, investigating the relationship between variables, e.g. from all records on COVID-19, we used association rule to determine which other keywords are related in research, like SARS, infection, virus, etc.

Apart from association rule, to design the three-fields plot in Fig. 12, we also applied clustering to discover and associate data records by countries of origin. The three-fields plot in Fig. 12, is similar to the research by country using the Web of Science result analysis tool, in Fig. 2. The difference in the visualisation is that in Fig. 12 we can see the keywords associations between data records from individual country. While in Fig. 2, we can only see classifications of data records by country. To find a regression function, that estimates the relationship between data records, with the smallest amount of error, we developed a collaboration network map (Fig. 13), using country in the network parameters, with equivalence normalisation, in a circle network layout, using Louvain clustering algorithm and the minimum number of edges set at 2.

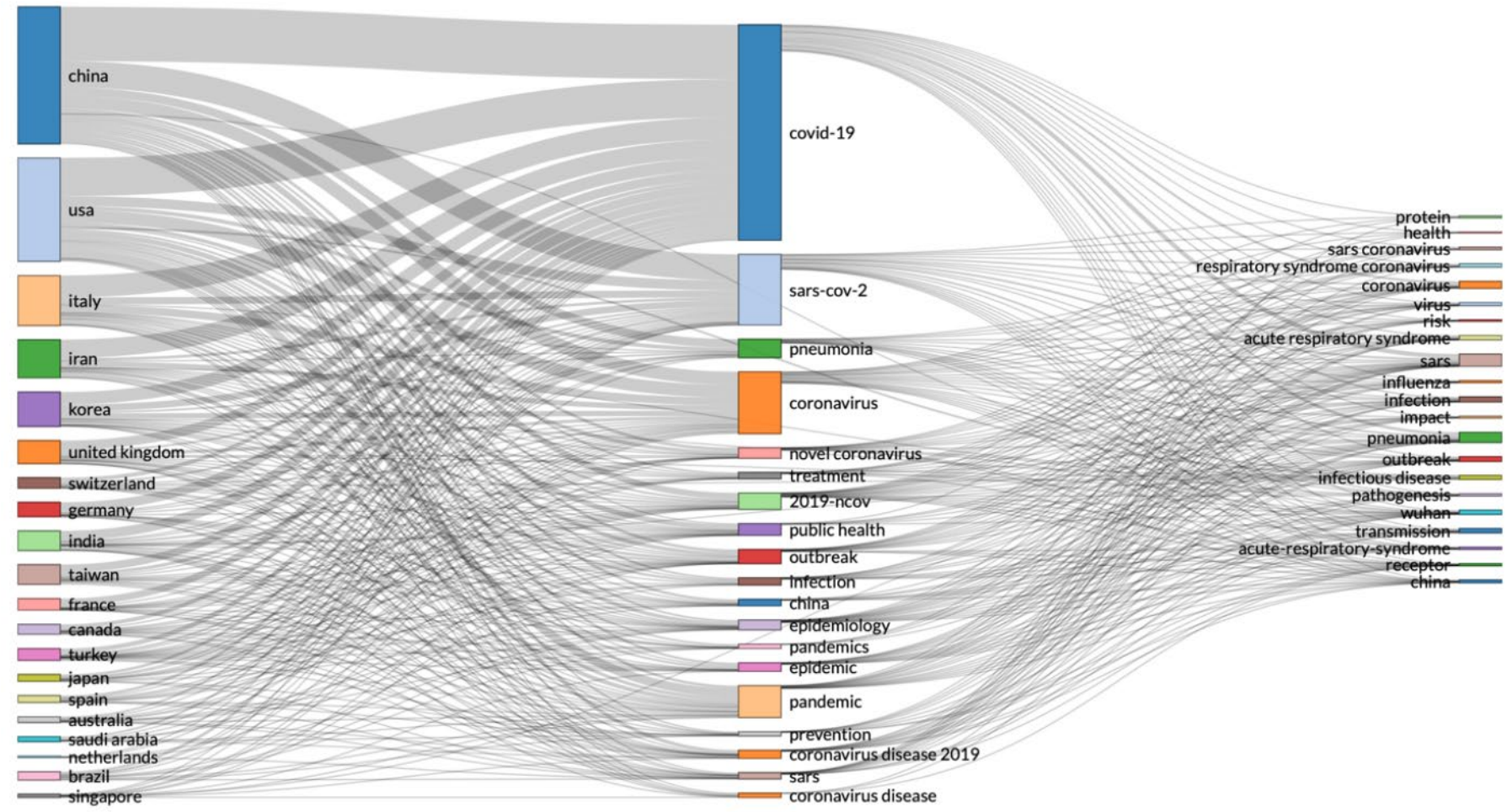

**Fig. 12** Three-fields plot of classification by country, research area and research keywords from all records on COVID-19

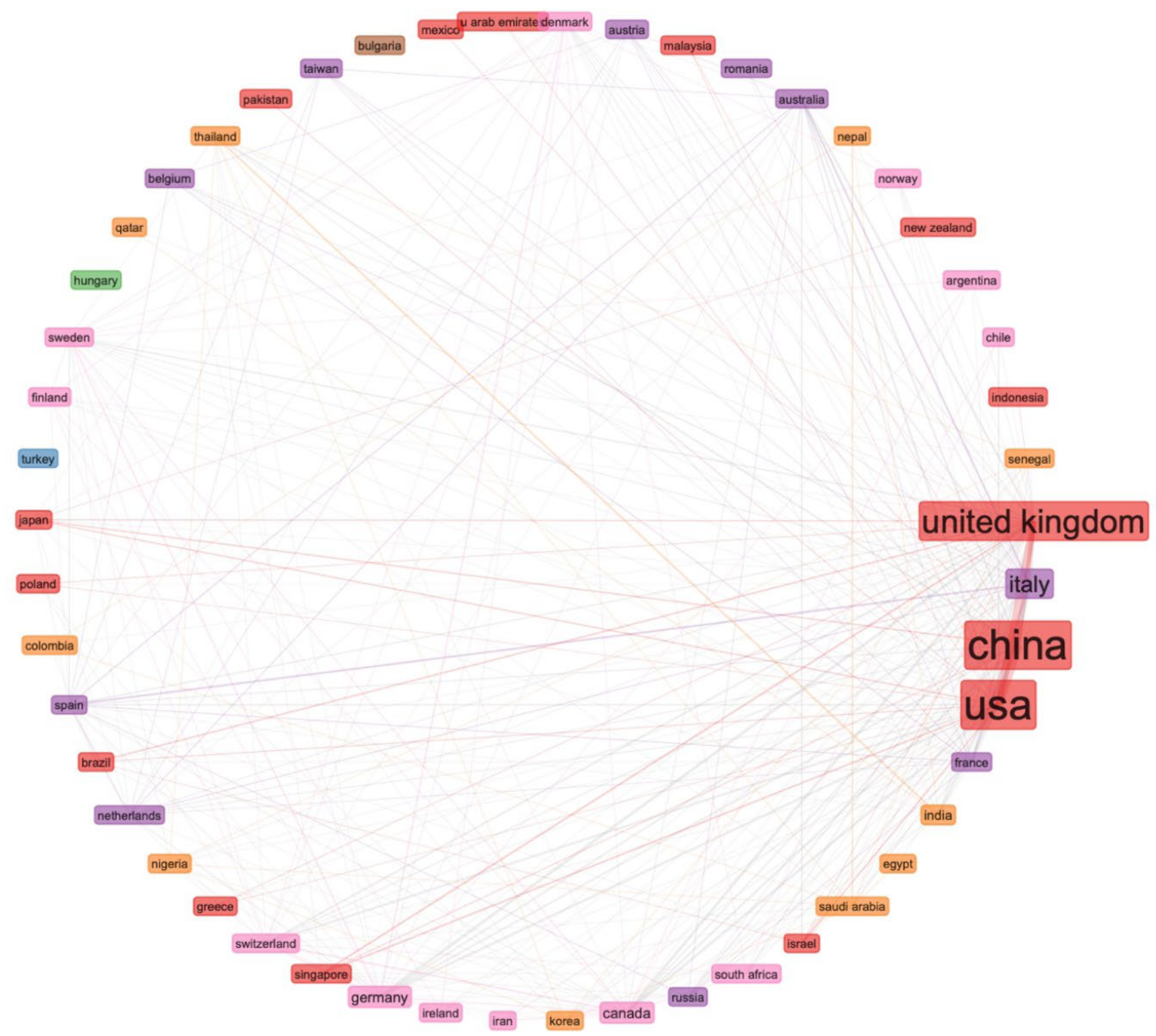

**Fig. 13** Collaboration network map by Country in a circle network layer with a minimum of 2 edges

Although very detailed, the collaboration network in Fig. 13 seems a bit cluttered. To present a better visualisation of the data records, in Fig. 14, we kept the same parameters, but we reduced the minimum number of edges at 7.

Since density is the proportion of present edges from all possible edges in the collaboration network, in Fig. 14, we can see the strongest collaboration networks, in edge connections and colour coding. Just to clarify these connections in the collaboration network map, edge density equals number of edges divided by maximal number of edges. Hence, an edge density in Fig. 14 is defined of overlapping and weighted in graph communities. However, it is possible that edge variations in multiple keywords mainly reflect the variations in few underlying keywords. Hence, in Fig. 15, we applied factorial analysis as a statistical method to identify joint keywords in response to unnoticed (concealed) keywords.

The parameters we applied in the factorial analysis (Fig. 15), included 'multi-correspondence analysis', with field of analysis being the keywords of the records, with automatics clustering and a maximum number of terms 50. In Fig. 15 we describe variability among the correlated keywords with potentially lower number of unobserved keywords (factors), aiming to identify independent latent keywords. In other words, we wanted to reduce the number of keywords in the data records. Our objective was to find the latent factors that create a commonality in the data records, and we applied factorial analysis because it is a statistical method that can identify smaller number of underlying variables, within large numbers of observed variables[2].

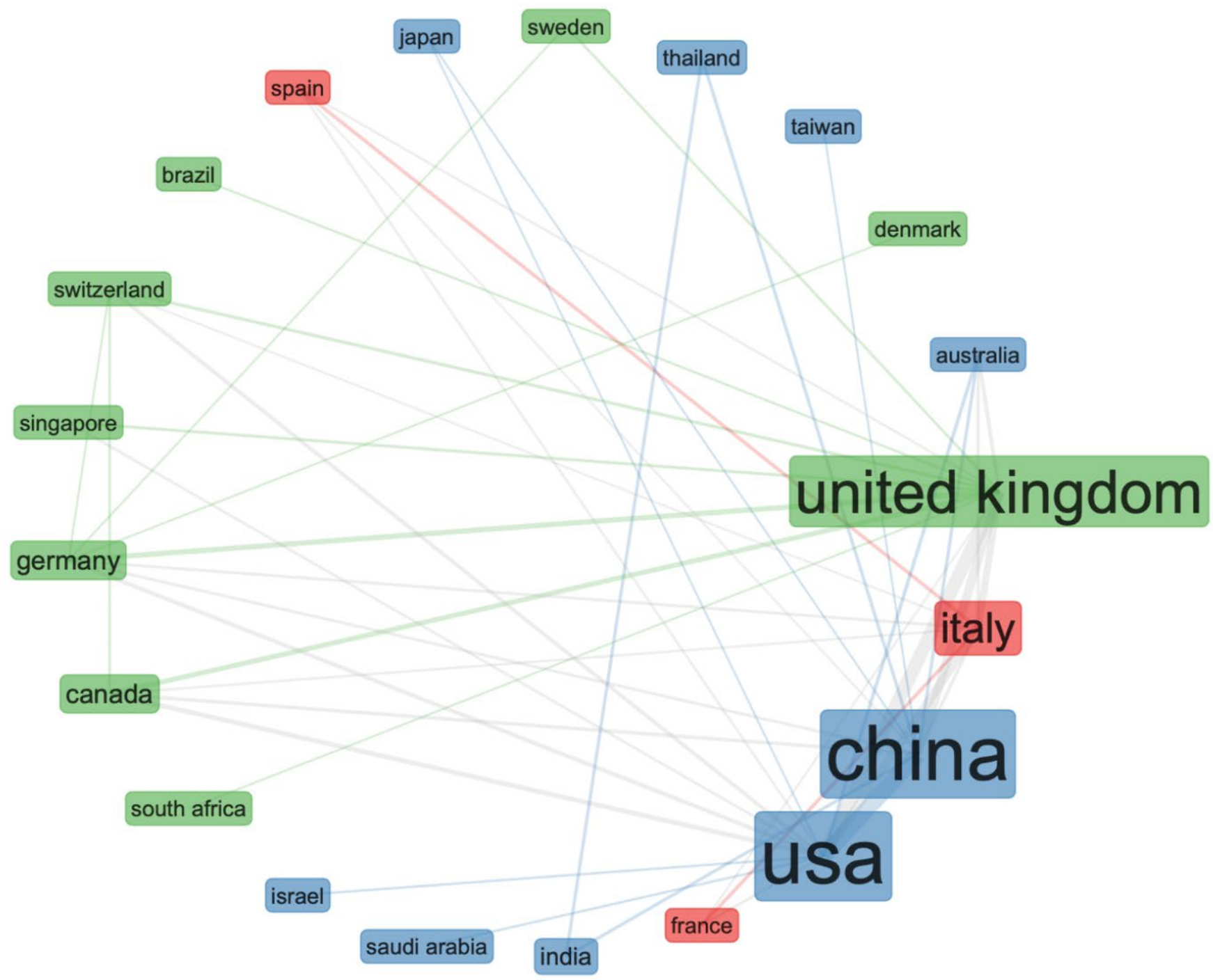

**Fig. 14** Collaboration network map by Country in a circle network layer with a minimum of 7 edges

The factorial analysis derives two classifications of keywords (in Fig. 15). The classification in blue, represents keywords like management, care, exposure, response, therapy, health, impact, risk, etc. The classification in red, represents more specific keywords, like respiratory syndrome, functional receptor, acute respiratory syndrome, etc. What we can see in the Fig. 15 conceptual structure map, is factorial analysis of 3094 records, presenting classification of common keywords from all data records, in two classifications.

## 5 Discussion

The most interesting findings from this study was that institutions that are established as leaders in scientific research on pandemic and epidemics, have responded much slower than organisations that are located in the areas where COVID-19 first occurred. Wuhan University is currently very high in the classification of scientific research on COVID-19. In the alternative classification, using historical records on most prominent organisations in this field, the Wuhan and Teheran Universities are not present in the statistical classification. This triggers many questions on have the leading organisations on pandemic and epidemic management reacted as the appropriate speed? Is so, why are they behind in the production of scientific journal? Has there been a gap in communications and data sharing? Leaving these organisations oblivious to what was happening? Or was it that our global alert mechanisms failed to act? Did we ignore the warning signs? These are just some of the emerging questions from this study. Until these questions are answered, conspiracy theories would continue to spread. With the COVID-19 pandemic slowing down, we should also be seeking answers to these questions to prevent a second wave, and most importantly, to prevent the same mistakes happening in future pandemics.

[2] https://en.wikipedia.org/wiki/Factor_analysis.

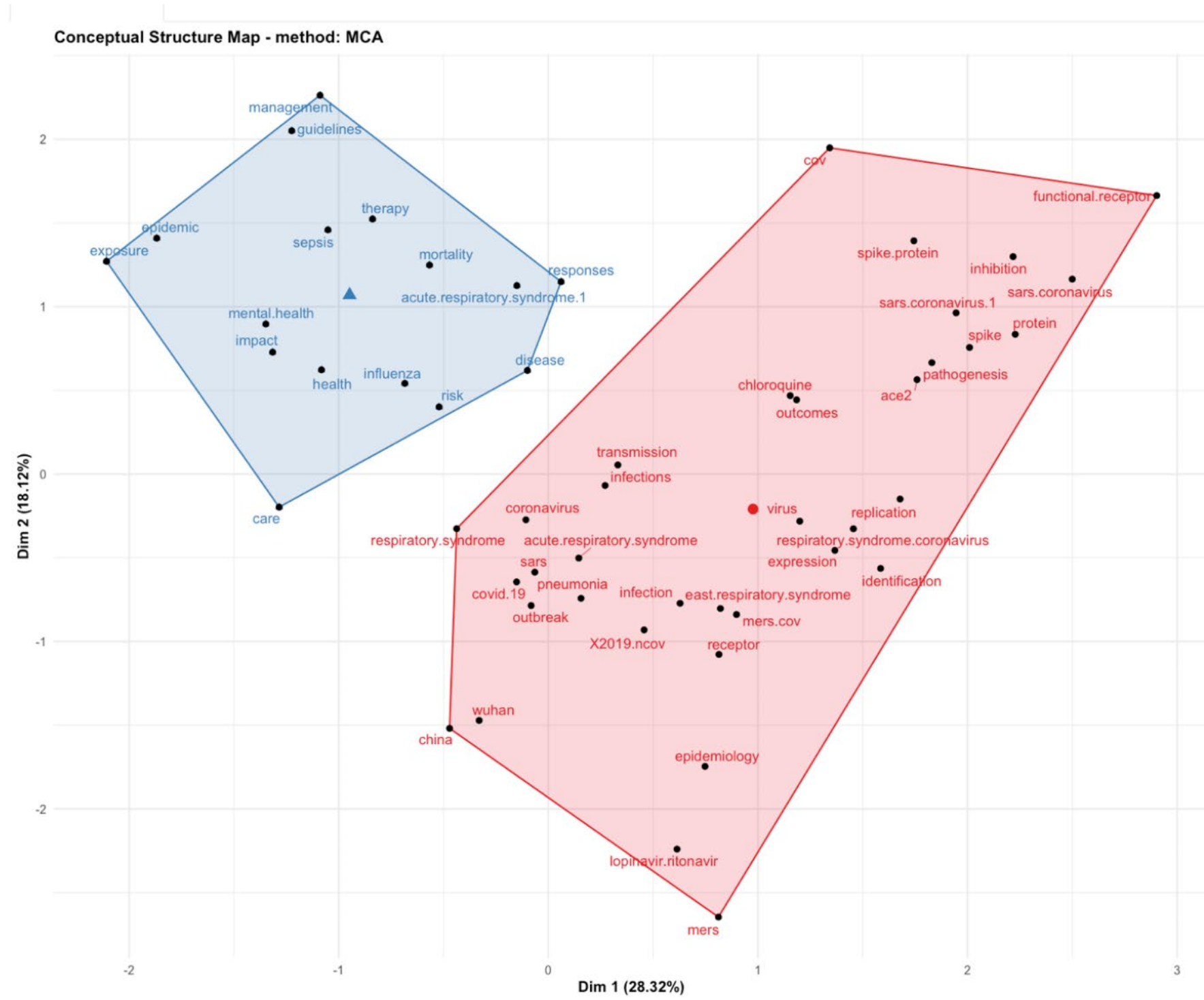

**Fig. 15** Factorial analysis - conceptual structure map

Our research findings can serve as the background work and starting point for future studies on developing spatial indices, such as large spatio-temporal datasets, and multidimensional objects, for computer decision support systems based on artificial intelligence.

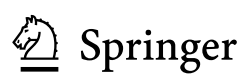

### 5.1 Discussion on results

As we can see from Fig. 2, the USA produced most scientific records on Covid-19 during the first wave, followed by China and the UK. In terms of best performing university – during the first wave, in Fig. 3 we can see that University of London was leading in the research efforts, followed by Harvard University and Huazhong University of Science Technology. However, it is worth mentioning that University of London (similarly to Harvard and Huazhong) is an umbrella organisation that represents many different universities. We tried to rectify this by separating the data by individual institutions in Fig. 4 and expanding the search to pandemics and epidemics. While the University of California systems emerged as the best performing university on a global level, the data is still partially representative of the umbrella organisations. We can see in Fig. 4 that University of London still appears on the list. Although Fig. 4 provides valuable insights on research by different institutions on pandemics and epidemics, we wanted to determine the best performing instruction on Covid-19 during the first wave, without the umbrella organisation. In Fig. 5, we managed to separate the data into individual organisations, and we can see that Huazhong University of Science Technology produced most research on Covid-19 during the first wave, followed by Wuhan University, Harvard Medical School, and University of Teheran Medical School. This changes the picture significantly from the analysis in Fig. 3. While it's difficult to confirm with certainty the connection between increased research output by individual institution, it is quite clear that the best performing institutions are based in countries / areas that were first impacted by the first Covid-19 wave. It could be that these institutions were best preforming, because of the urgency and the severity of the impact – in the 'snapshot in time' analysed. In the next step of our analysis, we wanted to compare this (first) postulate (we would need more date to call this a hypothesis) and we investigated if the same organisations would be expected to perform the best in an event of a global pandemic. In Fig. 6, we analysed scientific data records from 1900 to 2020 on the topic of pandemics and epidemics and not on Covid-19 specifically. The objective of this analysis was that grounded on the idea that the term 'Covid-19' was coined only after the pandemic occurred. In other words, this term (word) didn't exist before Covid-19 happened. Since this term didn't exist as a word, it should not be present in scientific data records prior to 2019 (the actual term/word was announced by WHO in 2020). In Fig. 6, we can see the analysis of the data records on pandemics and epidemics from 1900 to 2020, and it's quite clear that the organisations in Figs. 3 and 5 are not the same as the organisations in Fig. 6 (with the exception of Harvard University that preserved its second place). This supports the (first) postulate and confirms that the organisations that performed best, are not the organisations that have traditionally performed best in this field of research. The second postulate we present is that countries that got worst affected in the first wave, invested most money in research on Covid-19. This can be seen from Fig. 7, where the National Natural Science Foundation of China emerges as the largest funder of research on Covid-19. Worth mentioning that the data in Fig. 7 is categorised by organisation and not categorised by nation, and we can see that multiple organisations from China are in the top organisations that provided funding for Covid-19 research – during the first wave. This categorisation was done to compare the total research funding with organisations that are considered as largest funders in the more general field of pandemics and epidemics, which are analysed in Fig. 8. By comparing Fig. 7 with Fig. 8, we can clearly see that the leading organisations didn't allocate the most funding on Covid-19 during the first wave. This confirms the second postulate - that the worst affected countries in the first wave, invested most money in the initial research efforts on Covid-19. We understand that further research is required to prove these postulates as hypothesis. Hence, we are making our data records available (in open access) for future researchers to use the data sets that we collected as a 'snapshot in time' from the first wave of the Covid-19 pandemic. To eliminate bias in our analysis, we continued our analysis with different biometrical tools and software. We used the VOSviewer to present visualisations of the data records by country, with records mapping (in Fig. 9), by density (in Fig. 10), by collaborations (Fig. 11), with three-fields plot of classification by country, research area and research keywords (from all records on COVID-19) (Fig. 12), with a circle network of collaborations (Figs. 13 and 14), and with Factorial Analysis (Fig. 13).

## 6 Conclusions

As the scientific research on COVID-19 continues to expand, the publications are becoming more fragmented, which creates challenges in navigating through the accumulation of new knowledge - on global pandemics. In this article, we present the results from bibliometric science mapping based on three different data mining methods. The process can be replicated by other scientist seeking to analyse research records from the first response on the COVID-19 pandemic. We found individual tools being restrictive, and we propose a multi-tool approach that enables faster results from statistical and graphical packages, aligned to bibliographical databases. With the use of these statistical methods, we presented visualisations of the research connections between areas and countries, on the emerging patterns from national responses, and we provide scientific insights on the speed of response. Our aim was to provide statistical 'snapshot in time', and to assist other researchers to reassess the response in the initial stages of the pandemic and prepare for future global pandemics.

In the article, we presented two conclusions:

1. The best performing institutions are based in countries / areas that were first impacted (and most severely) by the first Covid-19 wave,
2. Countries that got worst affected in the first wave, invested most money in research on Covid-19 – during the first wave.

While there is significant evidence for these conclusions to be confirmed in this article, we believe this topic will be further investigated and analysed for many years to come. Hence, we make our datasets publicly available (in open access), for other researchers to reuse in future analysis.

There can be various interpretations in practice about these findings. The fact remains that the world was not prepared for a global pandemic. The research institutes that were expected to react as first responders, didn't respond as fast as the institutes and organisations in the most affected areas. In the end, we have seen that organisations that were preparing for a Disease X event, produced the most output. But during the first wave, most of the output was produced by organisations and institutes that had access to data on the Covid-19 pandemic. This brings into question the value of sharing medical data (at speed and low latency) in preventing and managing future Disease X events.

### 6.1 Research limitations

There are obvious limitations in interestingness metrics, such as lack of insights into negative relationships, lack of statistical base on COVID-19. In addition, since we can only present results that emerge from the data, this study lacks an objective criterion for assessment. By lack of objective criterion, we refer to the lack of clearly defined research objectives, in specific terms that can be used to confirm if the terms of the objective criterion definitions are met. We didn't have a predefined problem or a research question that we tried to answer, such as; is one country or organisation better than other. Instead, the visualisations in this article are representative of the statistical data records, as described in our search parameters, available on the 17th of May 2020. In the spirit of reproducible research, we include our data records in this submission.

**Supplementary information** The online version contains supplementary material available at https://doi.org/10.1007/s40745-022-00406-8.

**Acknowledgements** Eternal gratitude to the Fulbright Visiting Scholar Project.

**Author contribution** Dr Petar Radanliev: main author; Prof. Dave De Roure, Prof. Max Van Kleek: supervision; Rob Walton, Omar Santos, La'Treall Maddox: review and corrections.

**Funding** This work was funded by the UK EPSRC [grant number: EP/S035362/1] and by the Cisco Research Centre [grant number 1525381].

**Availability of data and materials** all data and materials included in the submission.

**Code Availability** N/A.

## Declarations

**Conflict of interest** On behalf of all authors, the corresponding author states that there is no conflict nor competing interest.